\documentclass[a4paper,aps,amsmath,amssymb,nopacs,10pt,twocolumn,floatfix]{revtex4}
\usepackage{graphicx}
\usepackage{graphics,color}
\usepackage{umlaut}

\begin{document}

\keywords{effective field theory, spin glass order, replica symmetry breaking, random interactions
%, transport effects.
}
\title[Low T spin glass order]{Construction and purpose of effective field theories\\
for frustrated magnetic order}
\author{R. Oppermann and M.J. Schmidt}
\affiliation{Institut f. Theoretische Physik, Universität Würzburg, Am Hubland, 97074 Würzburg, FRG}
\date{March 21, 2007}
\begin{abstract}
This article reviews recent years' progress in the low temperature analysis of standard models of spin glass order such as the Sherrington-Kirkpatrick (SK) model. Applications to CdTe/CdMnTe layered systems and explanation of glassy antiferromagnetic order at lowest temperatures stimulated us to study in detail the beautifully complex physical effects of replica symmetry breaking (RSB). We discuss analytical ideas based on highly precise numerical data which lead to the construction of relatively simple effective field theories for the SK model and help to understand the mysterious features of its exact solution. The goal is to find construction principles for the theory of interplay between frustrated magnetic order and various relevant physical degrees of freedom. The emphasis in this article is on the role of Parisi's RSB, which surprisingly creates critical phenomena in the low temperature limit despite the absence of a standard phase transition.
\end{abstract}
\maketitle
\section{Introduction}
In many fields of physics the discovery of an effective field theory represents a milestone of theoretical progress and sometimes allows analytical answers where otherwise only the brutal power of computers can find structure-less numerical hints. Let us mention nonlinear sigma models where nonlinearly realized symmetries \cite{bibJZJ} play a central role. Applications to slow dynamics in magnets, to the localizing effects of disorder in Anderson localization and, more recently, as a model for non-equilibrium dynamics of spin glasses \cite{bibchamonlfc} are well-known examples.

In the field of spin glasses, Talagrand \cite{bibtalagrand} proved recently that Parisi replica symmetry breaking \cite{bibparisi} represents, as advocated since three decades, the exact structure of the basic SK model solution. Explicit solutions, which are necessary for applications in solid state physics at low temperatures, were however not yet available except in the vicinity of the critical temperature.

Numerical methods combined with the renormalization group of De Dominicis \cite{bib3DzeroT},
aiming at a relation between Parisi- and droplet-picture \cite{bibdroplet},
appears to support the Parisi picture even for finite range interaction, and including low temperatures.

Phenomenological theories which study the interplay between frustrated magnetism with other degrees of freedom, for example with electronic transport, exist \cite{bibweissman}, but systematic approaches were limited to the region of the spin glass freezing temperature, or neglected the details of Parisi RSB. As the literature shows, most applications were, probably due to the complication of the Parisi solution and the way how to include it, concerned with precursor effects in the nonmagnetic region above the critical temperature \cite{bibdobros}. In the case of quantum phase transitions (driven by a non-thermal parameter) the nonmagnetic region can extend down to $T=0$. Since the specific features of the frustrated order are not involved (only through fluctuation-precursors of glassy order), no specific theoretical modeling such as Parisi RSB or the droplet picture appear in these dynamic mean field theories. The glassy order however by itself is a highly interesting physical object: being on one hand an equilibrium solution and on the other as a fluctuation solution in a new pseudo-time space \cite{bibprl2007} as we shall explain below.

Thus it seems clear that a theory of quantum spin glasses and its applications needs effective field theories which simplify the problem but maintain undamaged the essential physics. The basic difficulty is to find the minimal requirements for such theories such as symmetries, low energy excitations etcetera.
Work on the effective field theory for the standard model, called Sherrington-Kirkpatrick (SK) model, seemed indispensable. It led us to simplifications \cite{bibprl2007,bibprl2005}, which have the power to generate even analytical results where standard approaches cannot penetrate even initial stages. We discuss the low temperature analysis of RSB which serves to find key symmetries of spin glass order.
%
%
%---------------------------------------------------------------------------------
\section{Glassy antiferromagnets with sublattice-asymmetry and the Parisi-model of hierarchical frustrated magnetic order}
%---------------------------------------------------------------------------------
%
%
%---------------------------------------------------------------------------------
\subsection{\bf Application to dilute magnetic semiconductor layers in a field}
%---------------------------------------------------------------------------------
%
The 'true' spin glass character of CdMnTe has been questioned for many years. In order to give clear hints the theoretical challenge was to distinguish this phase from anti-ferromagnetic clusters. In our work on excitonic magnetic polarons in CdMnTe we explained a maximum of the polaron field-induced magnetization as a specific feature of the spin glass order \cite{bibcdyy}.
The theoretical description of CdMnTe requires two order parameters which are given by the standard magnetization $M_i=\left<S_i^a\right>$ and the by the glass order parameter
$$q^{ab}=\left<Q_i^{ab}\right>=\left<S^a_i S^b_i\right>,$$
where $i$ denotes lattice-sites while $a$ and $b$ stand for the (theoretical) replicas of the random magnet. The matrix field $Q$ measures the spin-overlap between the different replica-labeled disorder realizations. The Ising spins $S$ are Villain spins which do not represent the Mn-Heisenberg spins individually but describe rather the orientation of a tetrahedral complex of manganese spins (normalized Villain spins $S^a=\pm1$ imply diagonal q-elements to become $q^{aa}=1$).
%
%---------------------------------------------------------------------------------
\subsection{\bf Modulated replica symmetry breaking for asymmetric sublattices:\\
spin glass--ferrimagnetic--antiferrimagnetic double phase transition }
%---------------------------------------------------------------------------------
%
Motivated by the successful application to the spin glass phase of CdMnTe at moderate temperatures and in a confined magnetic field, a natural challenge was to provide solutions for the general phase diagram including the $T=0$ limit. Moreover, the antiferromagnetic interactions between the Mn-ions are expected to induce a phase transition into a glassy antiferromagnetic phase for high enough manganese concentration.
Clearly infinite-range interactions cannot distinguish sites and hence do not seem to be appropriate to describe antiferromagnetic order; however Korenblit and Shender created a way out of this dilemma by allowing infinite range 'inter-site' interactions, which act only between between spins on different sublattices of the antiferromagnet \cite{bibkorenblit}. In a series of papers they derived the $\infty$-RSB close to $T_c$ and also considered corrections from 'intra-site' interactions.

We generalized our Villain-Ising-spin model for CdMnTe by defining a Korenblit-Shender type model Hamiltonian \cite{bibosk} with field-generated asymmetry of sublattices $A$ and $B$
\begin{multline}
{\cal H}=-\sum\sum J_{i_A,j_B}S(r_{i_A})S(r_{j_B})+\\\sum h(r_{i_A})S(r_{i_A})+\sum h(r_{i_B})S(r_{i_B}),
\end{multline}
where $r_{i_A}$ denotes sites on sublattice $A$. The $J_{i_A,j_B}$ stand for the random Korenblit-Shender interaction, acting between sites of different sublattices, and are Gaussian-distributed around an antiferromagnetic mean value $J_{af}$ and fluctuation width $J$. Thus the model interpolates between the clean antiferro-limit $J=0$ and the broad distribution limit, where ferro- and antiferromagnetic interactions have equal weight.

The order parameter field $Q^{ab}$ not only bifurcates into independent matrix fields $Q_A$ and $Q_B$ for each sublattice, but, originating in the decoupling procedure of the disorder-generated 4-spin interaction \cite{bibkorenblit}, a third field denoted by  $Q_{AB}$ emerges with average $\left<Q_{AB}\right>=\left<Q_A\right>+\left<Q_B\right>$ which complicates the theory considerably.

The effective magnetic field $H_A$ acting on sites of sublattice $A$ reads in $1$-step RSB
\begin{equation}
%H_{\kappa}(h)=h-J_{af}M_{\bar{\kappa}+J\sqrt{q_{\bar{\kappa},2}}z_{\kappa,2}
%+J\sqrt{q_{\bar{\kappa},1}-q_{\bar{\kappa},2}}z_{\kappa,1}
H_{A}(h)=h-J_{af}M_{B}+J\sqrt{q_{_{B,2}}}\hspace{.1cm}z_{_{A,2}}+J\sqrt{q_{_{B,1}}-q_{_{B,2}}}\hspace{.1cm}z_{_{A,1}}
\end{equation}
and vice versa for sublattice $B$; this provides interesting hints on the structure of phase diagram and the possible phase transitions. A competition between homogeneous field $h$ and random magnetic interaction with dominating antiferromagnetic part leads first to a continuous phase transition (spin glass to dirty ferrimagnetic phase) with order parameter $M_A-M_B\neq0$ with both magnetizations pointing in the same direction, followed by a discontinuous second transition into an antiferrimagnetic phase with $sign(M_A)=-sign(M_B), |M_A|\neq |M_B|$.

In all phases the matrix $Q$ has nonvanishing mean values due to the randomness. In phase I $Q$ is of course the only order parameter (though a field-generated finite magnetization $M=M_A=M_B$ is present for all temperatures), but between phase II and III its role may be less important. In any case, both $A$-$B$-asymmetric phases allow to apply the modulated RSB scheme in comparison with standard Parisi RSB.
\begin{figure*}
\centering
\includegraphics[width=300pt]{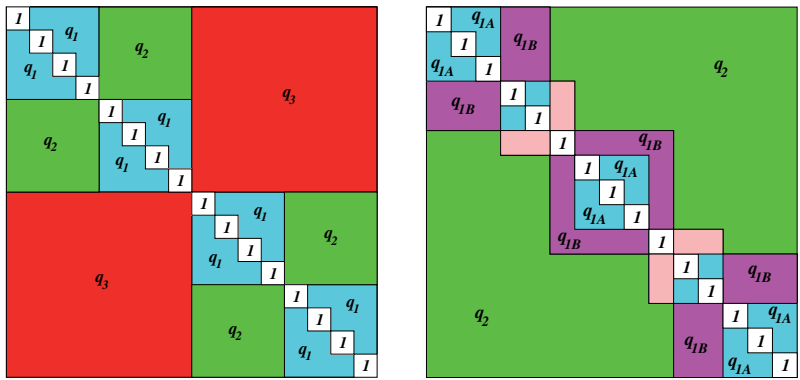}
\caption{Figure shows typical examples of a matrix in 2-step Parisi RSB(left side) and a modulated 1-step RSB example (right) according to our embedding scheme.}
\label{fig:q-matrices}
\end{figure*}

The stability of Parisi RSB had already been questioned in several directions: not only the alternative droplet picture but also perturbations of its suspected marginal stability were analyzed. In the case of CdMnTe the spin glass-antiferromagnetic transition at high Mn concentrations, we considered the A-B asymmetry as a potential perturbation of the RSB scheme itself \cite{bibosk}. Thus the application of RSB to a glassy antiferromagnet in a field such as CdMnTe \cite{bibosk} turns into an RSB-stability test with respect to Parisi block overlaps, ie Parisi block sizes (representing cluaters of pure states) which are different on sublattice A and B.

The standard scheme of RSB \cite{bibosk} uses the hierarchical embedding of smaller structures as shown in Fig.(\ref{fig:q-matrices},left).
The modified scheme for asymmetric sublattice sites on A and B allows all embedded objects to have different size
on A- and B-sites, which leads to the overlap of A- and B-matrices. Thus intersections occur as shown in the Fig.(\ref{fig:q-matrices},right) together with embedding of say smaller A-boxes within smaller B-boxes. Thus we found it convenient to introduce a new embedding parameter $\gamma$, which is not present in the standard RSB-scheme and which is to be determined by the selfconsistent procedure. Indeed we found nontrivial selfconsistent solutions. The energies for standard RSB and for modified RSB, calculated by using the corresponding selfconsistent solutions, turned out to be different and we observed a crossing at the discontinuous ferrimagnet-antiferrimagnetic glass transition at $J_{af}/J\approx0.4$ for a polaron field $h=4J$.

The modulated scheme was thus shown in $1$-step RSB to become the preferred solution with respect to standard RSB
in the glassy antiferrimagnet. However this does not yet prove its relative stability and the proposed transition between two different types of RSB. It is necessary to go beyond $1$-step RSB and to approach the $\infty$-RSB solution reliably. At that time, this wasn't even achieved for low temperatures in the standard Parisi scheme. We had to focus on this problem before the modulated RSB can be resolved in higher RSB orders. We cannot anticipate whether the $n$-th order modulated RSB can show more and more embedding parameters which eventually add a new dimension to the parameter-space(s). This remains a very interesting problem for the future.
%
%
%---------------------------------------------------------------------------------
\section{Analytic modeling of high precision numerical data for the standard spin glass model in the low temperature limit}
%---------------------------------------------------------------------------------
%
The search for the so-called equilibrium solution of the standard Sherrington-Kirkpatrick spin glass model and its fermionic generalization \cite{bibhubbardsk} has been an ambitious project even in the $T=0$-limit. The equilibrium solution is important in general, since it can distinguish between the Parisi picture of hierarchical order and the droplet picture \cite{bibdroplet}.
Only in the former picture spin glass order can exist in a magnetic field.

Finding the $T=0$-solution (or at least a good approximation) provides in general a useful anchor-point for the determination of phase diagrams. In more complicated models, where transitions between glassy phases with RSB exist, the solution of RSB in the SK-model is an indispensable first step. In order to achieve this goal we transformed the nested Parisi free energy for finite temperatures \cite{bibparisi} and the selfconsistent equations, which result form its extremization, into a $T=0$ energy with new variational parameters and new selfconsistent structures \cite{bibprl2005}. The original parameter set
$\{q_k,m_k\}$ (required to determine $Q$ at finite RSB) transforms into $\{q_k,a_k\}$ with $a_k\equiv lim_{T\rightarrow 0} m_k/T$ and a few more subtleties outlined in Ref.\cite{bibhfro}.
%
%---------------------------------------------------------------------------------
\subsection{\bf Renormalization group in RSB space:\\ decimating RSB-orders in analogy with Wilson's momentum shell integration}
%---------------------------------------------------------------------------------
%
The hierarchical structure of the Parisi scheme was either solved exactly in the continuum limit near the transition temperature or solved numerically in low orders exhibiting increasing failure due to essential singularities as $T\rightarrow0$.

\begin{figure*}
\centering
\includegraphics[width=500pt]{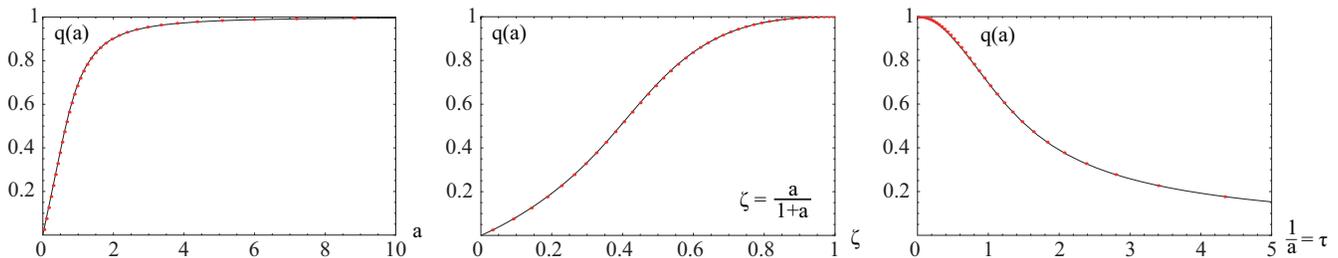}
\caption{The confluent hypergeometric model for the $T=0$ spin glass order parameter function
(continuous lines) is shown together with $40$-step RSB selfconsistent numerical data (dots). Three different plots demonstrate the good agreement as a function of $a$ (left), confined to a unit interval as a function of $a/(1+a)$ (middle), and as a function of the inverse $1/a$ which acquires the meaning of a pseudo-time $\tau$ (right).}
\label{fig:qfig}
\end{figure*}

In \cite{bibprl2005} we have demonstrated scaling behaviour by mapping the $T=0$-parameters $q_k$ and $a_k$, as obtained for the five lowest nontrivial RSB-orders, onto one curve, which together with the self-similar structures in the Parisi scheme suggested the existence of a renormalization group.
The decimation of RSB-orders and consecutive renormalization of the model-parameters was postulated in analogy with Wilson's renormalization group where momentum shell contributions were integrated out \cite{bibwilsonRG}. These model parameters were not only the discretized values of the order parameter $q_k$ (which becomes the order parameter Parisi function in the continuum limit), but also the Parisi block sizes $a_k$ which are associated with the clusters of pure magnetic states. For finite-order RSB, which we considered to be analogous to a finite-size system in real space, the numerical RG was defined by the fact that an order function model with only a small number of parameters (between $1$ and $4$) fitted well the high precision numerical data for almost all RSB-orders). The reason was that the
assumption of an underlying field theory allows, after each decimation) the rescaling to its original form and hence the retrieval of the identical order function with renormalized parameters.
The simplest order function model $q(a)=\frac{\sqrt{\pi}}{2}\frac{a}{\xi} erf(\frac{\xi}{\sqrt{a^2+w}})$ contained a correlation length parameter, which appeared to flow towards the small value $\xi\approx 1.13$ (together with $w\rightarrow 0$). The parameter $\xi$ sets also the scale for the RSB-orders \cite{bibprl2005}; thus the RSB-results should converge fast, the main changes occurring in 1st- and 2nd-order of RSB, which was confirmed by the flow of the linear susceptibility, of the single-valley susceptibility, and of the energy under increasing RSB-order \cite{bibprl2007}.
An improved model for $q(a)$ was found in \cite{bibprl2007} to be given by $q(a)=\frac{a}{\sqrt{a^2+w}}{_1}F_1(\alpha,\gamma,-\frac{\xi^2}{a^2+w}),$ where two additional parameters modeled well a nonlinear feature in the small-$a$ regime \cite{bibprl2007}. The simpler error-function model is contained for $\alpha=\frac12$ and $\gamma=\frac32$, since ${_1}F_1(\frac12,\frac32,-x^2)=\sqrt{\pi}erf(x)/(2x)$. The best parameter set for the full range $0<a<\infty$, matching also the extrema of $q'(a)$, was found to be $\{\alpha\approx 0.53,\gamma\approx 1.71, w\approx 0.02, \xi\approx 1.16\}$
\cite{bibprl2007}. We show a comparison with $40-$step RSB (80-dimensional extremization of the SK-energy) in Fig.(\ref{fig:qfig}).
\vspace{-1cm}
%
%
%---------------------------------------------------------------------------------
\subsection{\bf Mysterious Coulomb analogies of frustrated magnetic order}
%---------------------------------------------------------------------------------
%
The analytical modeling of numerical data several remarkable analogies between the $T=0$ SK spin glass and Coulomb systems \cite{bibprl2007}\\
1. spectra of Parisi block size ratios $r_k\equiv a_k/a_{k-1}$ for an infinite number of $a_k$ approaching zero and an equally infinite number of divergent $a_k$ in the $\infty-$RSB limit reminded of the bound-state energy spectrum of a non- or relativistic particle in a Coulomb field,\\
2. the internal field distribution $P(h)$ can be mapped (in the $\infty-$RSB limit) to the Efros-Shklovskii law for the density of states of strongly localized interacting electrons (in two dimensions), (it can as well be mapped to the density of states of the fermionic SK-model \cite{bibosk} in all orders of RSB)\\
3. the order function $q(a)$ is well represented by the so-called Coulomb wavefunction ${_1}F_1$.

It seems a matter of field theories to explain why such unexpected similarities between frustrated magnetic order and strongly localized electrons emerge.
We can however also draw some immediate interesting conclusions from point 1., which favour nonanalytic power law behaviour. Let us reconsider the Parisi block sizes $m_k(T)$ as a function of temperature and, knowing their linear low $T$-behaviour for almost all $k$, write their Taylor series in powers of temperature as
$$m_k(T)=a_k T+O(T^2).$$
For those $k$, which show either vanishing or diverging $T$-coefficients in the $\infty-$RSB limit, the $m_k$ cannot behave linearly in $T$ as usual. It is a natural possibility that the Taylor series breaks down and allows for nonanalytic $T$-behaviour.
This particular behaviour would then be associated with the unexpected discrete spectra of ratios $m_k/m_{k-1}$ which should equal the $a_k/a_{k-1}$ mentioned in point 1.

%
%---------------------------------------------------------------------------------
\section{\textbf{Effective field theory and a new order parameter function}}
%---------------------------------------------------------------------------------
%
The good agreement between the confluent hypergeometric function and the high-order numerical data is now used to propose a field theoretic model, which is much simpler to handle than the original spin model. One goal is to find a theory which can be required to satisfy independent criteria like a variational principle or symmetry considerations. The reproduction of numerical data would then serve as a test of the proposal, which may not be unique. It is not the only and even not the major goal to approach an exact analytical solution of the standard model - instead, approximations which allow generalizations to quantum spin glass phases interacting with coupling to charge degrees of freedom (electronic transport for example) represent even more attractive goals.

Let us now take advantage of the confluent hypergeometric function as a good analytical model for the SK order parameter at zero temperature. Since the small parameter $w$ decays monotonically in the regime of all calculated RSB-orders, we consider here $w=0$ in the continuum limit ($\infty$-RSB) for simplicity.

The well-known defining differential equation of ${_1}F_{1}$ (often called Coulomb wave-function) is easily transformed into the useful form \cite{bibprl2007}
\begin{equation}
\frac12 \frac{a^4}{\xi^2}\frac{d^2}{d a^2}q(a)
+\left(\left(\frac32-\gamma\right)\frac{a^2}{\xi^2}-1\right)a\frac{d}{d a} q(a)+2\alpha\hspace{.1cm} q(a)=0
\label{2ndorderdiffeq}
\end{equation}
This 2nd order differential equation should satisfy the boundary conditions $q(0)=0$ and $q(\infty)=1$. Since the numerical data are well fitted with $\gamma >3/2$, the 2nd linearly independent solution need not be considered. A physical interpretation becomes more obvious once we have transformed Eq.(\ref{2ndorderdiffeq}) into 1st-order differential equations. This is achieved by the defining two auxiliary functions \cite{bibgradryz}
\begin{equation}
p_1(a)=exp\left\{\frac{2}{\xi}\int\left(\left(\frac32-\gamma\right)\frac{\xi}{a}-\frac{\xi^3}{a^3}\right)da\right\},
\end{equation}
and
\begin{equation}
 p_2(a)=4\alpha\frac{\xi^4}{a^4}p_1(a)
\end{equation}
which result in the Riccati equation
\begin{equation}
z'(a)+\frac{1}{p_1(a)}z(a)^2+p_2(a)=0,\quad a\geq 0,
\end{equation}
where
\begin{equation}
z(a)\equiv p_1(a)q'(a)/q(a).
\end{equation}

It is convenient to introduce the new order parameter function $\phi(a)$ by the definition
\begin{equation}
\phi(a)\equiv \frac{1}{\sqrt{2\alpha}}\left(\frac{a}{\xi}\right)^{2\gamma-1}e^{-\xi^2/a^2}z(a)
\end{equation}
which obeys a Langevin-type equation
\begin{equation}
\frac{d}{d(\frac{\xi}{a})}\phi(a)=\frac{\delta}{\delta\phi(a)}{\cal H}[\phi]
\end{equation}
with
\begin{equation}
{\cal H}[\phi]=\sqrt{8\alpha}\hspace{.05cm}\left(\phi(a)+\frac16\phi(a)^3\right)
-\left((\gamma-\frac12)\frac{a}{\xi}+\frac{\xi}{a}\right)\phi^2(a)
\label{cubic-deq}
\end{equation}
where the field $\phi$ is related to the order function $q(a)$ by
\begin{equation}
\phi(a)=-\frac{1}{\sqrt{2\alpha}}\frac{d}{d\frac{\xi}{a}}log\left(q(a)\right).
\label{cubic-deq}
\end{equation}
In this form three different facts became obvious:\\
1. the differential equation has the appearance of a relaxational equation and gives rise to the interpretation of $1/a$ being a physical pseudo-time,\\
2. it can be viewed as the local limit of a Halperin-Hohenberg equation \cite{bibHH} of slow dynamics, and\\
3. the Hamiltonian can be recognized as the expansion of a $sinh$-form up to 3rd order in the order parameter field $\phi(a)$.\\

The first point relates the pseudo-relaxation as a feature of the so-called equilibrium Parisi solution at $T=0$ to the non-equilibrium dynamics \cite{biblfc}. We have given a particular interpretation in \cite{bibprl2007}.

The 2nd point concerns usually only critically slow dynamics at finite temperatures, where fast modes are integrated out contributing to or generating the leading relaxational behaviour and in addition lead to the presence of a stochastic field. We come back to this field later in a different context, since the SK model itself does not give rise to such a term. Instead the dynamical equation is a complete though approximative representation of the SK model.

The latter point strongly suggests to define and to analyze the related model with Hamiltonian
\begin{equation}
\frac{d}{d(\frac{\xi}{a})}\phi(a)=\frac{\delta}{\delta\phi(a)}{\cal H}[\phi]
\end{equation}
with
\begin{equation}
{\cal H}[\phi]=\sqrt{8\alpha}\hspace{.05cm}\sinh(\phi(a))-\left((\gamma-1/2)a/\xi+\xi/a\right)\phi^2(a)
\label{sinh-deq}
\end{equation}
\vspace{.3cm}
\begin{figure*}
\includegraphics[width=500pt]{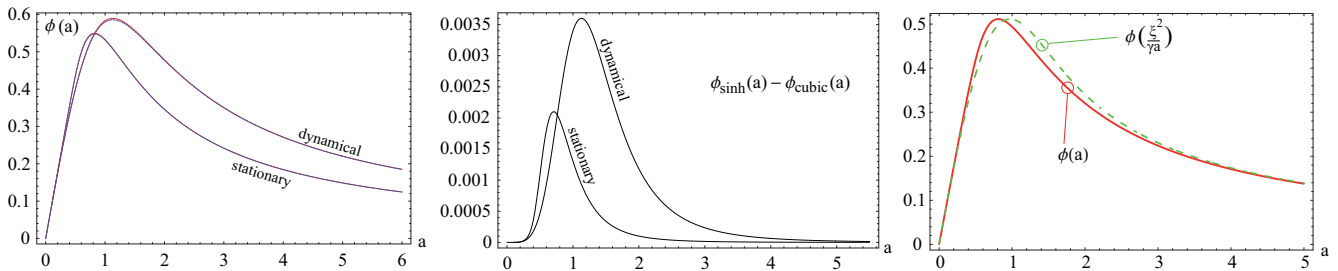}
\caption{Figure shows field $\phi$-solutions of the sinh-model and of the cubic model (dashed curves, almost coinciding solutions for both models) together with the solutions which extremize $\cal H$ (left Figure);
the small difference between these solutions is exhibited in the middle, while right Figure compares our model function $\phi(a)$ with $\phi(\frac{\xi^2}{\gamma a})$, which show identical asymptotic behaviour but deviate in the region of largest $\phi$.}
\label{fig:field-phi}
\end{figure*}
It turns out that the solutions of the cubic model and the numerical ones of the sinh-model are almost identical, as illustrated in Fig.\ref{fig:field-phi}.

In addition we notice that the pseudo-dynamical behaviour measured by the 1st order a-derivative (lhs) contributes little: if we attempt a solution by extremization of the energy density at each fixed pseudo-time $a$ (neglecting the lhs), the solution of the algebraic equation differs only in the regime of maximal $\phi$. Of course a physical justification for such a variational principle still has to be found.

In contrast to the order parameter function $q(a)$ which vanishes at the critical point $a=0$ (in zero field) and monotonically increases towards $q=1$ at the 2nd critical point $a=\infty$, the new function $\phi(a)$ vanishes also at the critical point $a=\infty$. Our proposal of an order parameter function interpolating between two critical limits is very unusual. It is linked to only one field theory interpolating between $a=0$ and $a=\infty$, while normally one would expect two distinct critical theories to apply. Our goal is however currently best achieved by one single theory, which allows the desired generalization to finite range interactions and (real-) time-dependent quantum dynamics.

Neglecting the small corrections from the pseudotime-derivatives of both models, one arrives at an extremization 'principle' for $\cal H$.

One observes an invariance of Eqs.(\ref{cubic-deq},\ref{sinh-deq}) under $a\rightarrow -a$ with $\phi(a)=-\phi(-a)$ and ${\cal H}$ both antisymmetric.
The asymptotic behaviour of its solutions in the small-$a$- and large-$a$-regime can also be mapped by
\begin{equation}
\frac{a}{\xi}\rightarrow {\frac{1}{\gamma}}\hspace{.1cm}\frac{\xi}{a}
\end{equation}
as Fig.(\ref{fig:field-phi}) (right) shows. It remains to be seen whether the asymptotic behaviour of $\phi(a)$
which determines the asymptotic behaviour of $q(a)$ as well, can be related to the different discrete spectra found for the Parisi block size ratios \cite{bibprl2007}.

In this chapter we have presented a possible field theoretic origin of the analytic order function model $q(a)=\frac{a}{\sqrt{a^2+w}}{_1}F_1(\alpha,\gamma,-\frac{\xi^2}{a^2+w})$; the spreading parameter $w$ was set equal to zero as suggested by the numerical RG-flow of the data-matching parameters $\{\alpha,\gamma,w,\xi\}$. It is not yet proved that $w^*=0$ is an attractive fixed point;  in case that $w^*$ would remain small but finite (as favoured by extrapolation of the numerical data), one would hardly see any difference in the order function $q(a)$ and $\phi(a)$ (while $q'(a)$ would show a cutoff-effect with $q'(0)$ finite).
Research in the future must show to which extent the confluent hypergeometric order function can be improved, while keeping its relatively simple form for useful applications of spin glass order effects in general solid state theory.
%
%---------------------------------------------------------------------------------
\section{Definition of toy models of Halperin-Hohenberg type for double-time spin glass dynamics and outlook}
%---------------------------------------------------------------------------------
%
Cubic field theories in terms of the matrix order parameter field $Q$ are very common in spin glass physics, be that the classical theory or a quantum version. In these examples a fourth-order term in $Q$ has been required to generate a replica symmetry broken solution. Let us recall the quantum phase transition theory of metallic spin glasses of Ref.\cite{bibsro}, which dealt with fluctuations in real space and real time (disregarding replica symmetry breaking as usual in dynamic mean field theories in the ordered phase.
This cubic field theory turned out to be a nontrivial quantum extension of Michael Fisher's renormalization theory\cite{bibsro} of the classical Yang-Lee edge singularity.

In the preceding section we have by contrast described a cubic field theory for replica symmetry breaking itself.

The relaxational form with a single time-derivative usually results from integrating out fast modes which exchange energies with the slow modes that one wishes to describe explicitly. A known technique to derive such time-reversal breaking (for the subset of slowly varying fields) from an initially time-reversal invariant quantum system is the Mori-Zwanzig formalism. The procedure leads also to the appearance of a stochastic field.
It does not exist in our SK-model representation, but could be added to eventually account for the coupling of our spin system to fast modes and in general to other degrees of freedom. In the quantum case this would immediately imply also quantum-dynamical behavior depending on the real time t. Under these conditions we would propose a model of Halperin-Hohenberg type \cite{bibHH} (a type of model that also occurs in nonequilibrium models of classical and quantum problems \cite{biblfc})
\begin{equation}
(\partial_{\frac1a} +\kappa \hspace{.1cm}\partial_t) \phi(a,t)= \frac{\delta {\cal H}}{\delta\phi(a,t)}+\zeta(a,t)
\end{equation}
with gaussian white noise
\begin{equation}
\left<\zeta(a,t)\zeta(a',t')\right>= \Gamma\hspace{.08cm}\delta(t-t')\delta(a-a'),
\end{equation}
and in case of finite range effects in d-dimensional real space
\begin{equation}
(\partial_{\frac1a} +\kappa \hspace{.03cm}\partial_t) \phi(a,x,t)= \frac{\delta {\cal H}}{\delta\phi(a,x,t)}+\zeta(a,t),
\end{equation}
\begin{equation}
\left<\zeta(a,x,t)\zeta(a',x',t')\right>= \Gamma\hspace{.02cm}\delta(t-t')\delta(x-x')\delta(a-a').
\end{equation}
The constant $\Gamma$ in the stochastic field correlations will in general be temperature dependent but does not decay to zero in the $T=0$-limit. The stochastic field should represent the effect of inelastic interactions in quantum models (energy transfer with fast modes, which are integrated out), for example electronic noise acting on extremely slow spin dynamics. The SK-model is contained as the local limit of the Halperin-Hohenberg type equations \cite{bibprl2007}. The remaining single variable $1/a$ recalls the 'time' used by Sompolinsky to represent the Parisi solution as a dynamical one \cite{bibsompol}.

The generalization to quantum problems will require a modified Hamiltonian $\cal H$ and the perhaps modified second differential equation relating $\phi(a)$ and $q(a)$. A double-time dynamic mean field theory for quantum models with range-free interactions in real space like the Heisenberg quantum spin glass \cite{bibqhsg} is another field of application, since numerical methods can soon provide enough data for comparison.

Replica-free methods like Schwinger-Keldysh \cite{bibkeldysh} or supersymmetry techniques \cite{bibJZJ} should be applied, in order to support the search for the characteristic symmetries of the double criticality of the SK-model, and the analysis of other models must be intensified.

{\bf Acknowledgements.} We thank the SFB410 of the DFG for support.

\end{document}